\documentstyle[twoside,psfig]{article}

\catcode`\@=11
\long\def\@makefntext#1{
\protect\noindent \hbox to 3.2pt {\hskip-.9pt
$^{{\eightrm\@thefnmark}}$\hfil}#1\hfill}               

\def\@makefnmark{\hbox to 0pt{$^{\@thefnmark}$\hss}}    

\def\ps@myheadings{\let\@mkboth\@gobbletwo
\def\@oddhead{\hbox{}
\rightmark\hfil\eightrm\thepage}
\def\@oddfoot{}\def\@evenhead{\eightrm\thepage\hfil
\leftmark\hbox{}}\def\@evenfoot{}
\def\sectionmark##1{}\def\subsectionmark##1{}}

\oddsidemargin=\evensidemargin
\newcounter{sectionc}\newcounter{subsectionc}\newcounter{subsubsectionc}
\renewcommand{\section}[1] {\vspace{12pt}\addtocounter{sectionc}{1}
\setcounter{subsectionc}{0}\setcounter{subsubsectionc}{0}\noindent
        {\tenbf\thesectionc. #1}\par\vspace{5pt}}
\renewcommand{\subsection}[1] {\vspace{12pt}\addtocounter{subsectionc}{1}
        \setcounter{subsubsectionc}{0}\noindent
        {\bf\thesectionc.\thesubsectionc. {\kern1pt \bfit #1}}\par\vspace{5pt}}
\renewcommand{\subsubsection}[1] {\vspace{12pt}\addtocounter{subsubsectionc}{1}
        \noindent{\tenrm\thesectionc.\thesubsectionc.\thesubsubsectionc.
        {\kern1pt \tenit #1}}\par\vspace{5pt}}
\newcommand{\nonumsection}[1] {\vspace{12pt}\noindent{\tenbf #1}
        \par\vspace{5pt}}

\newcounter{appendixc}
\newcounter{subappendixc}[appendixc]
\newcounter{subsubappendixc}[subappendixc]
\renewcommand{\thesubappendixc}{\Alph{appendixc}.\arabic{subappendixc}}
\renewcommand{\thesubsubappendixc}
        {\Alph{appendixc}.\arabic{subappendixc}.\arabic{subsubappendixc}}

\renewcommand{\appendix}[1] {\vspace{12pt}
        \refstepcounter{appendixc}
        \setcounter{figure}{0}
        \setcounter{table}{0}
        \setcounter{lemma}{0}
        \setcounter{theorem}{0}
        \setcounter{corollary}{0}
        \setcounter{definition}{0}
        \setcounter{equation}{0}
        \renewcommand{\thefigure}{\Alph{appendixc}.\arabic{figure}}
        \renewcommand{\thetable}{\Alph{appendixc}.\arabic{table}}
        \renewcommand{\theappendixc}{\Alph{appendixc}}
        \renewcommand{\thelemma}{\Alph{appendixc}.\arabic{lemma}}
        \renewcommand{\thetheorem}{\Alph{appendixc}.\arabic{theorem}}
        \renewcommand{\thedefinition}{\Alph{appendixc}.\arabic{definition}}
        \renewcommand{\thecorollary}{\Alph{appendixc}.\arabic{corollary}}
        \renewcommand{\theequation}{\Alph{appendixc}.\arabic{equation}}
        \noindent{\tenbf Appendix \theappendixc #1}\par\vspace{5pt}}
\newcommand{\subappendix}[1] {\vspace{12pt}
        \refstepcounter{subappendixc}
        \noindent{\bf Appendix \thesubappendixc. {\kern1pt \bfit #1}}
        \par\vspace{5pt}}
\newcommand{\subsubappendix}[1] {\vspace{12pt}
        \refstepcounter{subsubappendixc}
        \noindent{\rm Appendix \thesubsubappendixc. {\kern1pt \tenit #1}}
        \par\vspace{5pt}}

\newcommand{\textlineskip}{\baselineskip=13pt}
\newcommand{\smalllineskip}{\baselineskip=10pt}

\def\eightcirc{
\begin{picture}(0,0)
\put(4.4,1.8){\circle{6.5}}
\end{picture}}
\def\eightcopyright{\eightcirc\kern2.7pt\hbox{\eightrm c}}

\newcommand{\publisher}[2]{{\begin{center}\footnotesize\smalllineskip
        Received #1\\
        Revised #2
        \end{center}
        }}

\def\abstracts#1#2#3{{
        \centering{\begin{minipage}{4.5in}\footnotesize\baselineskip=10pt
        \parindent=0pt #1\par
        \parindent=15pt #2\par
        \parindent=15pt #3
        \end{minipage}}\par}}

\newcommand{\bibit}{\nineit}

\renewenvironment{thebibliography}[1]
        {\frenchspacing
         \ninerm\baselineskip=11pt
         \begin{list}{\arabic{enumi}.}
        {\usecounter{enumi}\setlength{\parsep}{0pt}
         \setlength{\leftmargin 12.7pt}{\rightmargin 0pt} 
         \setlength{\itemsep}{0pt} \settowidth
        {\labelwidth}{#1.}\sloppy}}{\end{list}}

\newcounter{itemlistc}
\newcounter{romanlistc}
\newcounter{alphlistc}
\newcounter{arabiclistc}

\newcommand{\fcaption}[1]{
        \refstepcounter{figure}
        \setbox\@tempboxa = \hbox{\footnotesize Fig.~\thefigure. #1}
        \ifdim \wd\@tempboxa > 5in
           {\begin{center}
        \parbox{5in}{\footnotesize\smalllineskip Fig.~\thefigure. #1}
            \end{center}}
        \else
             {\begin{center}
             {\footnotesize Fig.~\thefigure. #1}
              \end{center}}
        \fi}

\newcommand{\tcaption}[1]{
        \refstepcounter{table}
        \setbox\@tempboxa = \hbox{\footnotesize Table~\thetable. #1}
        \ifdim \wd\@tempboxa > 5in
           {\begin{center}
        \parbox{5in}{\footnotesize\smalllineskip Table~\thetable. #1}
            \end{center}}
        \else
             {\begin{center}
             {\footnotesize Table~\thetable. #1}
              \end{center}}
        \fi}

\def\@citex[#1]#2{\if@filesw\immediate\write\@auxout
        {\string\citation{#2}}\fi
\def\@citea{}\@cite{\@for\@citeb:=#2\do
        {\@citea\def\@citea{,}\@ifundefined
        {b@\@citeb}{{\bf ?}\@warning
        {Citation `\@citeb' on page \thepage \space undefined}}
        {\csname b@\@citeb\endcsname}}}{#1}}

\newif\if@cghi
\def\cite{\@cghitrue\@ifnextchar [{\@tempswatrue
        \@citex}{\@tempswafalse\@citex[]}}
\def\citelow{\@cghifalse\@ifnextchar [{\@tempswatrue
        \@citex}{\@tempswafalse\@citex[]}}
\def\@cite#1#2{{$\null^{#1}$\if@tempswa\typeout
        {IJCGA warning: optional citation argument
        ignored: `#2'} \fi}}

\def\fnt#1#2{\footnotetext{\kern-.3em
        {$^{\mbox{\scriptsize #1}}$}{#2}}}

\def\fpage#1{\begingroup
\voffset=.3in
\thispagestyle{empty}\begin{table}[b]\centerline{\footnotesize #1}
        \end{table}\endgroup}

\def\runninghead#1#2{\pagestyle{myheadings}
\markboth{{\protect\footnotesize\it{\quad #1}}\hfill}
{\hfill{\protect\footnotesize\it{#2\quad}}}}
\font\tenrm=cmr10
\font\tenit=cmti10
\font\tenbf=cmbx10
\font\bfit=cmbxti10 at 10pt
\font\ninerm=cmr9
\font\nineit=cmti9

\font\eightrm=cmr8

\def\qed{\hbox{${\vcenter{\vbox{                        
   \hrule height 0.4pt\hbox{\vrule width 0.4pt height 6pt
   \kern5pt\vrule width 0.4pt}\hrule height 0.4pt}}}$}}

\begin{document}
\setlength{\textheight}{7.7truein}  

\runninghead{$~$}{$~$}

\normalsize\textlineskip
\thispagestyle{empty}
\setcounter{page}{1}


\vspace*{0.88truein}

\fpage{1}
\centerline{\bf DOUBLY-SPECIAL RELATIVITY:}
\smallskip
\centerline{\bf FIRST RESULTS AND KEY OPEN PROBLEMS}
\vspace*{0.37truein}
\centerline{\footnotesize GIOVANNI AMELINO-CAMELIA}
\baselineskip=12pt
\centerline{\footnotesize\it Dipartimento di Fisica,
Universit\'{a} di Roma ``La Sapienza'', P.le Moro 2}
\baselineskip=10pt
\centerline{\footnotesize\it 00185 Roma, Italy}

\vspace*{0.225truein}

\publisher{(received date)}{(revised date)}

\vspace*{0.21truein}
\abstracts{I examine the results obtained so far
in exploring the recent proposal
of theories of the relativistic transformations between inertial observers
that involve both an observer-independent velocity scale
and an observer-independent length/momentum scale.
I also discuss what appear to be the key open issues for
this research line.}{}{}



\section{Introduction}
Over the last two years there has been a significant research effort
aimed at the development of ``Doubly-Special-Relativity"
(or ``DSR") theories~\cite{dsr1dsr2},
with contributions from about a dozen research groups~$^{1-19}$.
These theories of the relativistic transformations between inertial observers
involve both an observer-independent large-velocity scale
and an observer-independent small-length/large-momentum scale.
Some of the DSR results can already be seen as robust, and
provide encouragement for the idea that DSR might plausibly
have a role in quantum-gravity/quantum-spacetime research,
which was my main motivation~\cite{dsr1dsr2}
in proposing the DSR research programme.
As we start to establish some characteristic features of the DSR framework,
we also start realizing that there are a few recurring themes of DSR
research, issues that several authors have attempted to address,
but still await a satisfactory analysis.

Here I intend to review the key results obtained so far
in exploring the DSR framework, and I intend to discuss
the key open issues for this research line.
I will also present some preliminary
ideas which might be useful in research concerning
these open issues.

\section{Doubly Special Relativity}
In Galilei Relativity there is no observer-independent scale,
and in fact (for example) the dispersion relation is written
as $E=p^2/(2m)$ (whose structure fulfills the requirements
of dimensional analysis without the need for dimensionful
coefficients). As experimental evidence in favour of Maxwell equations
started to grow, the fact that those equations involve a
fundamental velocity scale appeared to require (assuming the Galilei
symmetry group should remain unaffected) the introduction
of a preferred class of inertial observers (the ``ether").
Einstein's Special Relativity introduced the first observer-independent
relativistic scale (the velocity scale $c$), its dispersion relation
takes the form $E^2 = c^2 p^2 + c^4 m^2$ (in which $c$ plays a crucial
role for what concerns dimensional analysis), and the presence
of $c$ in Maxwell's equations is now understood not as a manifestation
of the existence of a preferred class of inertial observers but
as a manifestation of the necessity to deform the Galilei
transformations. The Galilei transformations would not leave invariant
the relation $E^2 = c^2 p^2 + c^4 m^2$, which is instead an invariant
according to the Lorentz transformations
(the Lorentz transformations are a dimensionful
deformation of the Galilei transformations).

I argued in Refs.~\cite{dsr1dsr2} that it is not unplausible that
we might be presently confronted with an analogous scenario.
Research in quantum gravity\footnote{I am here thinking of some
preliminary results in the study of
Lie-algebra noncommutative spacetimes~\cite{kpoinap,kpoinPLB97},
the study of particle propagation~\cite{aemn1,grbgac}
in the ``Liouville-Noncritical-String"~\cite{emn}
model of spacetime foam,
the study of loop-quantum-gravity scenarios~\cite{gampul,mexdisprel,thiemdisprel},
and the study of critical superstrings
in an external B-field background~\cite{stringreviewnc}.}
is increasingly providing reasons of interest
in deformed dispersion relations
of the general type $c^4 m^2 =E^2 -  c^2 \vec{p}^2 + f(E,\vec{p}^2;E_p)$,
where $E_p \sim 10^{28} eV$ is the Planck scale.
Interest in these deformed dispersion relations is also coming as
a result of the realization that they may provide a solution
for the emerging GZK cosmic-ray anomaly (see below).
The fact that these dispersion relations involve an absolute energy
scale, $E_p$, was leading to the assumption that
a preferred class of inertial observers might be introduced.

Similarly, deformed dispersion relations had also emerged in
the mathematics programme of ``quantum deformations" of
classical algebras and groups. In particular,
deformed dispersion relations had emerged
in the study of the so-called $\kappa$-Poincar\'{e} Hopf algebras.
Within some of these Hopf algebras
there had been analyses of the action of exponentials of the
Lorentz-like generatos, and on the basis of the
results of these analyses
(also see Section~8)
it had been conjectured~\cite{rueggnew} that the actions described
by exponentials of the $\kappa$-Poincar\'{e} Lorentz-like generators
could not be combined to form a genuine group of transformations.
This again suggested that deformed dispersion relations could
not be implemented as genuine invariants of a (possibly quantum)
group of Lorentz transformations.

In Refs.~\cite{dsr1dsr2} I showed that it is incorrect
to assume that a deformed dispersion relation necessarily
implies the loss of covariance among inertial observers and
the emergence of a preferred class of inertial observers.
This observation has been then explored in detail in several
studies~$^{2-18}$.
Assuming for the sake of argument
that a deformed dispersion relation
was one day verified experimentally,
we now know that this observation would not automatically imply
that there is a preferred class of inertial observers.
At least for certain choices of the function $f$
in a dispersion relation
of type $ c^4 m^2 =E^2 -  c^2 \vec{p}^2 + f(E,\vec{p}^2;E_p)$
it is possible to give to the deformed dispersion relation
the same status that Special Relativity attributes to the
dispersion relation $ c^4 m^2 =E^2 -  c^2 \vec{p}^2$.
This however requires the introduction of
one more (besides $c$) observer independent
scale, $E_p$, and the structure of the
boost transformations must be modified in
such a way to leave
invariant the deformed dispersion relation.

Concerning the choice of the deformed dispersion relation
there appears to be at present a large level of uncertainty.
To describe the situation by an analogy one should imagine
attempts to introduce the first observer-independent
relativistic scale, a velocity scale ``$c$",
without the clear experimental indication that $c$ should
be the speed of massless particles and the maximum attainable
speed for massive particles.
As discussed later in this note, there are some experimental contexts
which appear to invite us to consider a second observer-independent scale,
some sort of high-energy scale, but these indications are only ``tentative"
and anyway they do not appear to single out a specific DSR theory.

In light of this situation, in proposing the DSR idea
I found appropriate~\cite{dsr1dsr2}
to at least discuss an illustrative example of this new kind
of relativistic theory. This example, now sometimes called ``DSR1",
has invariant dispersion relation
\begin{equation}
2 E_p^2 \left[ \cosh ({E \over E_p}) - \cosh ({m \over E_p}) \right]
= \vec{p}^2 e^{E/E_p}
~.
\label{dispKpoin}
\end{equation}
The DSR1 transformation rules, discussed in detail
in Refs.~\cite{dsr1dsr2,jurekrossano,judesvisser},
are very effectively characterized
through the amount
of rapidity needed to take a particle from its rest frame to a frame
in which its energy is $E$ (and its momentum is $p(E)$, which is
fixed, once $E$ is known, using the dispersion relation and the direction
of the boost)
\begin{equation}\label{xidsr1}
\cosh (\xi) = \frac{e^{E/E_p} - \cosh\left(m/E_p\right)}
  {\sinh\left( m/E_p \right)} ~,~~~
\sinh (\xi) = \frac{p e^{E/E_p}}
  {E_p \sinh\left(m/E_p\right)}\,\, .
\end{equation}

Another much studied example of DSR theory, which was proposed
more recently by
Maguejio and Smolin~\cite{leedsr} and is sometimes called ``DSR2",
has dispersion relation
\begin{equation}
{m^2 \over (1 - m/E_p)^2} = {E^2 - p^2 \over (1 - E/E_p)^2}
~,
\label{displee}
\end{equation}
and it prescribes that the amount
of rapidity needed to take a particle from its rest frame to a frame
in which its energy is $E$ should be given by
\begin{equation}\label{xidsr2}
\cosh (\xi) = \frac{E (1 - m/E_p)}{m(1 - E/E_p)} ~,~~~
\sinh (\xi) = \frac{p (1 - m/E_p)}{m(1 - E/E_p)} \,\, .
\end{equation}

Other DSR theories are also being developed, but these first two
illustrative examples remain the most studied ones.

\section{What is DSR? What is not DSR?}
Unfortunately, it is not uncommon to find in the physics literature
a rather sloppy use of terms such as ``fundamental scale".
In particular, ``fundamental scales" are often discussed as if they
were all naturally described within a single category.
For DSR research it is instead rather important that these concepts
be introduced very carefully.

\subsection{$c$-type versus $\hbar$-type fundamental scales}
The DSR framework intends to introduce (at least) one more
observer-independent relativistic scale.
The prototype of relativistic scale is of course $c$.
In the formulas reported in the preceding section the scale $E_p$
is exactly on the same footing as $c$.
But not all fundamental scales are introduced that way.
A good counter-example is the quantum-mechanics scale $\hbar$.

Space-rotation symmetry is a classical continuous
symmetry.
One might, at first sight, be skeptical that
some laws (quantum-mechanics laws) that discretize angular momentum
could enjoy the continuous rotation symmetry,
but more careful reasoning~\cite{areaNEWpap}
will quickly lead to the conclusion that there is no {\it a priori}
contradiction between discretization and a continuous symmetry.
In fact, the type of
discretization of angular momentum
which emerges in ordinary non-relativistic quantum mechanics
is fully consistent with classical space-rotation symmetry.
All the measurements that quantum mechanics still allows
(a subset of the measurements allowed in classical mechanics)
are still subject to the rules imposed by rotation symmetry.
Certain measurements that are allowed in classical mechanics
are no longer allowed in quantum mechanics, but of course those measurements
cannot be used to characterize rotation symmetry (they are not
measurements in which rotation symmetry fails, they are just
measurements which cannot be done).

Just as in classical mechanics,
when an observer $O$ measures the square-modulus  $L^2$
of the angular momentum, everything
can be said about how that square-modulus
appears to a second observer:
the value of the modulus is the same
for both observers. It happens to be the case that the values of $L^2$
are constrained by quantum mechanics on a discrete spectrum,
but this of course does not represent an obstruction
for the action of the continuous symmetry
on invariants, such as $L^2$.
When $O$ measures {\underline{exclusively}}
the $x$ component, $L_x$, of the angular momentum
it is not possible to predict the value of
any of the components
of that angular momentum along the $(x',y',z')$ axes of $O'$.
This is true at the quantum level just as much as it is
true at the classical
level. This is another example of situation in which
the fact that quantum mechanics constrains
the values of an observable, $L_x$, on a discrete spectrum is irrelevant
for rotation symmetry, since the relevant symmetry
does not prescribe how that same observable is seen by another
observer\footnote{Note that if another mechanical theory, clearly different
from quantum mechanics, allowed simultaneous eigenstates
of ${\hat{L}}_x$, ${\hat{L}}_y$, ${\hat{L}}_z$
and predicted discrete spectra for all of them,
then the classical continuous space-rotation symmetry would
inevitably fail to apply.}.

A more detailed discussion of this point can be found
in Ref.~\cite{areaNEWpap}. Essentially one finds that $\hbar$ is
not a scale pertaining to the structure of the rotation
transformations. The rotation transformations can be described
without any reference to the scale $\hbar$. The scale $\hbar$
sets, for example, the minimum non-zero value of angular
momentum ($L^2_{min}=3 \hbar^2/4$), but this is done in a way
that does not require modification of the action of rotation
transformations.

Galilei boosts are instead genuinely inconsistent with the
introduction of $c$ as observer-independent speed of massless
particles (and maximum velocity attainable by massive particles).
Lorentz transformations are
genuinely different from Galilei transformations.

Both $\hbar$ and $c$ are fundamental scales that establish
properties of the results of the measurements of certain observables.
In particular, $\hbar$ sets the minimum non-zero value of angular
momentum and $c$ sets the maximum value of speed.
But $\hbar$ has no role in the structure of the
transformation rules between observers,
whereas the structure of the transformation rules between observers
is affected by $c$.
I am describing $c$ as a {\underline{relativistic}} fundamental
scale, whereas $\hbar$ is a fundamental scale that does not affect
the transformation rules between observers.

A characterizing feature of the DSR proposal is that there should
be more than one scale playing a role analogous to $c$.
In particular, the illustrative examples discussed above,
DSR1 and DSR2, are relativistic theories with two observer-independent
relativistic scales, the ``speed-of-light" velocity scale $c$ and
Planck energy scale $E_p$.

One can try to introduce the Planck scale in analogy with $\hbar$
rather than with $c$. In particular, Snyder looked~\cite{snyder}
for a theory in which spacetime coordinates would not commute,
but Lorentz transformations would remain unmodified by the
new commutation relations attributed to the coordinates.
Such a scenario would of course {\underline{not}} be a DSR.

In closing this subsection I should comment on one more type
of fundamental constants. As I just explained, $\hbar$ and $c$
are different types of fundamental constants but they both
establish
properties of the results of the measurements of certain observables.
A third type of fundamental constants are the ``coupling constants".
For example, in our present description of physics the gravitational
coupling $G$ is a fundamental constant. It does not impose constraints
on the measurements of a specific observable, but it governs the laws of
dynamics for certain combinations of observables.
Also $G$ is observer indepedent, although a careful analysis (which
goes beyond the scopes of this note) is needed to fully characterize
this third type of fundamental scale. One can define $G$ operatively
through the measurement of static force between planets. In modern
language this amounts to stating that we could define $G$ operatively
as the low-energy limit of the gravitational coupling constant.
All observers would find the same value for this (dimensionful!)
constant.

This last remark on the nature of the fundamental constant $G$
is particularly important for DSR theories. In our present
description of physics the Planck scale is just the square root
of the inverse of $G$ rescaled through $\hbar$ and $c$.
The idea of changing the status of $G$ ({\it i.e.} $E_p$) from the one
of fundamental coupling scale to the one of relativistic
fundamental scale should have deep implications~\cite{dsr1dsr2,dsrpolon}.

In light of the preliminary success in contructing logically consistent
DSR theories, one can imagine one day to devise a ``triply-special"
relativity in which also the status of $\hbar$ is changed to
the one of relativistic fundamental scale.

\subsection{How many relativistic fundamental scales?}
Even within the DSR literature there has been some confusion about
the concept of relativistic fundamental scale,
and even on the counting of the relativistic fundamental scales
present in a given DSR theory. From the equations
(\ref{dispKpoin}),(\ref{xidsr1}),(\ref{displee}),(\ref{xidsr2})
it is clear that the theory has (at least in the energy-momentum
sector, to which all robust results have been so far confined)
two and only two relativistic fundamental scales.

The confusion has sometimes come from the fact that we are
used to thinking interchangeably of energy and inverse length.
One can of course think of the DSR deformation as a deformation
involving the Planck length $L_p$ rather than the Planck
energy scale $E_p$, but this of course must (implicitly or
explicitly) be understood in the sense that all energy-momentum
space is described in terms of inverse lengths/times.
Equivalently one can think of a DSR $L_p$-deformation as something
which characterizes wavelength/frequency space and the laws
of transformation for wavelengths and frequencies.

Instead some authors have taken literally/naively some DSR deformations
as  $L_p$-deformations of energy-momentum space, and they felt the
need to introduce a third scale (perhaps identifiable with $\hbar$)
to render dimensionless the product $L_p E$, between the Planck
length and the energy of a particle being studied relativistically.
This assumption is incorrect in the DSR theories discussed so far.
There is no need for a third scale and if one naively introduced it
by hand, in the way I just described, the scale would anyway not
be a genuine relativistic scale, since it would not genuinely
affect the structure of the laws of transformation between observers.

\subsection{Focus on the high-energy (high-momentum) regime}
Another main feature of the DSR proposal~\cite{dsr1dsr2}
is the objective of describing the short-distance/high-momentum
regime. Just like Galilei Relativity becomes inadequate
when velocities
are high (comparable to the speed-of-light scale),
the DSR research programme intends to explore the possibility that
Einstein's Special Relativity might itself become inadequate in
contexts in which ultrashort distances and/or ultrahigh
single-particle momenta are involved.
DSR's second relativistic observer-independent scale
should become relevant only in
short-distance/high-momentum regimes.

\subsection{Examples of the key role of the transformation rules:
dispersion relation, maximum momentum,
minimum-wavelength, and the number of relativistic scales}
As already mentioned and discussed in the later sections
of this note, there are some characteristic features that are
common to most (if not all) DSR theories.
Among these features much attention is directed in particular
to the emergence of a deformed dispersion relation and
the emergence of a maximum momentum/energy.
One might be tempted to characterize a theory as DSR if it leads
to these predictions, but actually a more careful attitude is
necessary.

Of course the fact that a quantum gravity scenario predicts a
Planck-scale deformation of the
dispersion relation does not automatically imply that
the laws of transformation between inertial observers be modified.
In a theory that pefectly obeys the laws of ordinary
Special Relativity one can of course find deformed dispersion relations
in contexts in which there is some sort of background, a medium,
a field distribution. And it is conceivable~\cite{grbgac}
that the structure of
spacetime at short distances, often heuristically described in
terms of a ``spacetime foam", might be described as a dispersion-inducing
background.

Similarly the concept of a maximum momentum (and other similar concepts)
does not automatically imply that
the laws of transformation between inertial observers be modified.
Rather than a reflection of deformed kinematics a maximum momentum
could for example be a reflection of the structure of the laws
of dynamics. Such a scenario has for example been explored
in the context of string-bits models.

In some DSR approached one finds a maximu momentum and adopts a
stadard relation between momentum and wavelength. As a result
the maximum momentum corresponds to a minimum wavelength.
Again this should be sharply distinguished from other
quantum-gravity scenarios~\cite{kempmang,dadebro}
in which one adopts a modified relation between momentum and wavelength,
such that, although momentum is still allowed to go up to infinity
(with boost transformations
possibly described exactly as in ordinary Special Relativity),
there is a finite minimum value of wavelength.

So, the issue of whether or not a given quantum-gravity scenario
requires the DSR framework cannot be addressed at the level of the
effects predicted within the physics-world picture of a single
given observer: it requires an investigation of the transformation
rules between inertial observers.

The transformation rules also define operatively the concept of a DSR
theory. DSR does not intend to modify the operative concept of
energy/momentum. We have adopted some ``energy-momentum measuring
devises" as the devises that provide an operative definition
of energy-momentum. Procedures suitable for such a direct operative
definition are for example in use at CERN and other particle-physics
laboratory. If we put these CERN devises on a spaceship and compare
the energy-momentum measurements reported by these spaceship devises
to the same results obtained by identical devises placed
on another spaceship, we presently predict that the measurement results
on the two spaceships be connected by Special-Relativity transformations
(in which the relative velocity of the spaceships place a key role,
which is fully specified by Special Relativity).
This comparison of measurement results is a well-defined operative
procedure for wich Special Relativity makes definite predictions
and DSR theories will make alternative predictions. The issue
can therefore at least in principle (and, as I will emphasize later,
in some cases even in practice) be settled experimentally.

The fact that one can introduce formal nonlinear maps between
different DSR theories has led some authors~\cite{jurekDSRnew}
into the erroneous conclusion that these theories might be equivalent.
Of course, the existence of
some nonlinear maps that connect different
DSR theories, does not imply the physical equivalence of the
theories: on the contrary it simply establishes the relations
between the {\underline{different}} physical predictions of
different DSR theories.
A possible source of this confusion may come from the fact that
those working in the DSR framework often have a formal General-Relativity
background. In General Relativity even a nonlinear map between
spacetimes may lead to physically equivalent spacetimes {\underline{if
corresponding changes are implemented on the metric tensor}}.
This is the core ingredient of the diffeomorphism invariance of
General Relativity. Spacetime coordinates are not themselves observable.
Distances (a concept which necessarily involves the metric tensor)
between spacetime point are observable, but the spacetime coordinates
are not themselves observable.
But these nonlinear maps that connect DSR theories are maps between
energy-momentum spaces, and energy/momentum are directly observable.
There is no diffeomorphism invariance of energy-momentum space.
Different DSR theories are therefore clearly inequivalent, and in fact,
as discussed in greater detail later in these notes,
they actually lead to different predictions for certain classes of
experiments, so we have identified physical contexts in which these
different theories give rise to explicitly different physical predictions
(and this of course excludes the physical equivalence of the theories).

\section{Doubly-Special-Relativity theories: DSR1, DSR2 and DSR3}
Up to this point I have explicitly described only the first two
illustrative examples of DSR theories, DSR1 and DSR2.
Each of these two examples is actually a good representative of
a corresponding class of DSR theories, which I will propose
to call DSR1-type theories and DSR2-type theories.

In introducing these classes of DSR theories I want to make
reference to their key physical predictions.
There has been extensive discussion in the recent quantum-gravity
literature of the fact that there are two classes of observations
which have extremely high sensitivity to possible Planck-scale deformations
of kinematics. The first context that deserves mention is the one
of experiments looking for possible Planck-scale-induced
wavelength-dependent relative delays in the times of arrival of
(nearly-) simultaneously emitted photons. In certain astrophysical
contexts~\cite{grbgac,biller,glast} this effect can be investigated
with very high sensitivity.
A second relevant context is the one of the threshold conditions
for particle production in certain collision processes.
When one of the colliding particles has ultrahigh energy and the
other particle is a very soft photon (a situation which is relevant
in certain astrophysical contexts, such as the GZK limit for
cosmic-ray physics), certain
Planck-scale deformations of special-relativistic kinematics
lead to effects that are observably large, particularly through their
implications~\cite{kifune,sato,gactp}
for the observations of ultra-high-energy cosmic rays.

The DSR1 theory is a theory in which, for particles with $E\ll E_p$,
there is indeed a wavelength dependence of the speed of photons,
and this effect comes in at the $E/E_p$ level,
as the reader can easily verify
applying the relation $v=dE/dp$ to the DSR1 dispersion relation.
This wavelength dependence at the level $E/E_p$ can be tested
with forthcoming experiments~\cite{grbgac,biller,glast}.
Instead DSR1 does not predict~\cite{dsr1dsr2,dsrpolon} an observably large
modification of the threshold conditions
for particle production in collision processes.

In the DSR2 theory the dispersion relation for photons ($m=0$)
is unmodified. There is thefore no wavelength dependence of the
speed of photons. In addition also the DSR2 theory
does not predict~\cite{frandar} an observably large
modification of the threshold conditions
for particle production in collision processes.

A DSR1-type theory will be any DSR theory in which the
wavelength dependence of the speed of photons comes in
at the level $E/E_p$, and there is no observably-large
modification of the threshold conditions
for particle production in collision processes.

A DSR2-type theory will be any DSR theory in which the
wavelength dependence of the speed of photons comes in
at the level $(E/E_p)^\alpha$, with $\alpha \ge 2$ (in particular
DSR2 has $\alpha = \infty$) and there is no observably-large
modification of the threshold conditions
for particle production in collision processes.

I recently proposed~\cite{dsrgzk} a third type of DSR theory,
DSR3-type theories, which is based on the presence of a specific
structure in the deformation of Special Relativity that DSR implements.
In ordinary Special Relativity
the amount
of rapidity needed to take a particle from its rest frame to a frame
in which it has energy $E$ is $\cosh (\xi) = E/m$.
In DSR1 and DSR2 this relation is modified
(see Eqs.~(\ref{xidsr1}),(\ref{xidsr2}))
and it is noticeable that the modifications
are structured in terms of the quantities $E/E_p$ and $m/E_p$,
which appear separately in the deformation.
I Ref.~\cite{dsrgzk} I argued that interest may be deserved
by DSR theories in which this deformation attributes a key role
to the quantity $E^2/(m E_p)$.
As an illustrative
example of the way in which the ``$\cosh (\xi)$ relation"
could be modified I considered the case
\begin{equation}\label{dispgood}
\cosh (\xi) = {E \over m}
(2 \pi)^{-E^2 \tanh[m^2 E_p^4/E^6]/(m E_p+E^2)} \,\, .
\end{equation}
The reader should not be deterred by the apparently {\it ad hoc}
form of Eq.~(\ref{dispgood}). We clearly do not yet have a compelling
DSR3-type theory, but relation (\ref{dispgood}) can be used to illustrate
some of the compelling features that these theories can have.

A DSR3-type theory will be a theory in which
the  ``$\cosh (\xi)$ relation" is modified in a way that attributes a key
role to the quantity $E^2/(m E_p)$.
DSR3-type theories may or may not have observably-large wavelength
dependence of the speed of photons (this depends on the full structure
of the theory, which is not specified by just giving
the ``$\cosh (\xi)$ relation").
DSR3-type theories will be characterized by important effects
in the regime in which the energy of the particle is in the neighborhood
of the value $\sqrt{m E_p}$. This may be particularly significant
in the study of modifications of the threshold conditions
for particle production in collision processes.
In particular the derivation of the GZK limit involves a proton
of energy $\sim 10^{19} eV$, and for a proton
the quantity $\sqrt{m E_p}$ takes the value $\sim 3 {\cdot} 10^{18} eV$.
Therefore the implications for the GZK limit are naturally
going to be observably large in DSR3-type theories.

\section{Brief tutorial on the construction of DSR theories}
Various strategies have been
exploited in constructing ``DSR theories"
(DSR-type laws of transformation
between inertial observers).
In formulating the DSR proposal
I thought~\cite{dsr1dsr2} it would be safest
to take a starting point ``{\it a la} Einstein":
introducing some postulates that were directly associated
with the relativistic observer-independent scales
and deriving from those postulates the structure of the
theory. I believe that this remains the most ``physical"
strategy, but other strategies appear to lead to equally-robust results.

While a careful examination of the DSR literature is of course
the best way to become familiar with relevant techniques,
here I want to describe briefly some lines of analysis that
apply to (what appears to be) the simplest class of DSR theories:
theories based on a deformed dispersion relation and
in which a key role is played by a nonlinear realization
of the Lorentz symmetry group.
I take as illustrative example the DSR1 theory.

In constructing DSR1 one can start from the
dispersion relation (which could, for example, be
the result of an experimental analysis)
\begin{equation}
2 E_p^2 \left[ \cosh ({E \over E_p}) - \cosh ({m \over E_p}) \right]
= \vec{p}^2 e^{E/E_p}
~.
\label{dispKpoinnew}
\end{equation}
This dispersion relation is clearly an invariant of space rotations,
but it is not an invariant of ordinary boost transformations.
The next step is the one of finding deformed boost
transformations which have (\ref{dispKpoinnew}) as an invariant.
In ordinary Special Relativity the boosts can be described by
\begin{equation}
B_a = i p_a {\partial \over \partial E}
+ i E {\partial \over \partial p_a}
~,
\label{boosnormal}
\end{equation}
and in the DSR1 case one can make the ansatz
\begin{equation}
{\cal B}_a = i \Delta_1(E,p^2,E_p) p_a {\partial \over \partial E}
+ i \Delta_2(E,p^2,E_p) {\partial \over \partial p_a}
+ i\Delta_3(E,p^2,E_p) p_a  p_b {\partial \over \partial p_b}
~,
\label{boosdelta}
\end{equation}
which is already aiming at automatically preserving space-rotation
symmetry.

Demanding that (\ref{dispKpoinnew}) is an invariant
of ${\cal B}_a$ transformations already provides a first condition
on $\Delta_1$, $\Delta_2$ and $\Delta_3$.
Next one should investigate the commutators of the ${\cal B}_a$
generators with the undeformed space-rotation generators
\begin{equation}
R_a = - i \epsilon_{abc} p_b {\partial \over \partial p_c} ~.
\label{rotnormal}
\end{equation}
The six generators ${\cal B}_a$, $R_a$
should close an algebra, and after some thinking one
can easily conclude that this algebra must still be
the usual Lorentz algebra. In fact, the operators  ${\cal B}_a$
and  $R_a$ have the same dimensions (in the sense of
elementary dimensional analysis) and this forbids the
introduction of the deformation scale $E_p$ in the
commutation relations that define the algebra.
One could perhaps consider an extension of the Lorentz algebra:
replacing the six-generator Lorentz algebra with a larger algebra.
For example, one could postulate commutators of the rotation/boost
generators that depend also on momenta. But this would
result in a theory in which finite rotation/boost transformations
do not really form a group: the way in which combinations of
rotation/boost transformations act on a given momentum would depend
on the value of that momentum, rather than being an intrinsic
property of the given combination of rotation/boost transformations.
[For example,
in ordinary Special Relativity the combination of two boosts
can give a rotation, and this rotation has no dependence
on the momentum on which the two boosts are applied.
This can be shown to be directly connected with the fundamental
concept of spacetime symmetry.]
It appears reasonable to reject\footnote{There is
a strong connection between this argument for rejecting deformations
of the Lorentz algebra and the problems which had been encountered
in the  $\kappa$-Poincar\'{e} literature in relation with the puzzling
emergence of a ''quasi-group" structure (see Section~9).}$~$~this type
of pathology of rotation/boost transformations,
and this leads to the only option of maintaining
the Lorentz algebra.
In spite of the deformation of the ${\cal B}_a$ generators,
the six generators ${\cal B}_a$, $R_a$
must close the usual Lorentz algebra. This leads to
other conditions on the $\Delta_1$, $\Delta_2$ and $\Delta_3$.

Combining the conditions coming from the invariance of
the dispersion relation and the Lorentz-algebra conditions
one can determine (up to some arguments of ''simplicity":
$\Delta_1$, $\Delta_2$ and $\Delta_3$
are not~\cite{dariotesi} fully fixed by the conditions, but
they are indeed fixed if they are assumed to have relatively
simple mathematical structure)
the full structure of the deformed boost generators.
In the DSR1 case one finds:
\begin{equation}
{\cal B}_a = i p_a \frac{\partial}{\partial E}+
i \left(\frac{1}{2E_p} {\vec p}^2
+E_p \frac{1-e^{-2E/E_p}}{2}\right ) \frac{\partial}{\partial
p_a}-i {p_a \over E_p} \left(p_b \frac{\partial}{\partial p_b} \right )~.
\label{dsr1boosts}
\end{equation}

It is useful to obtain explicit formulas
for the finite boost transformations that relate
the observations of two observers.
These are obtained by integrating the
familiar differential equations
\begin{equation}
{dE \over d\xi} = i [{\cal B}_a,E]
~,~~~{dp_b \over d\xi} = i [{\cal B}_a,p_b]
~,
\label{infp}
\end{equation}
which relate the variations of energy-moomentum with rapidity ($\xi$)
to the commutators between the boost generator and energy-momentum
(and of course these commutators
are implicitly coded in (\ref{dsr1boosts})).

In spite of the richer structure of the deformed boost generators,
the derivation of finite transformations from the structure
of the generators of infinitesimal transformations is rather
straightforward~\cite{dsr1dsr2,jurekrossano}, and can be done
in full generality.
Among the results that provide most insight in the structure
of the DSR theory are the mentioned ones that
specify the amount of rapidity needed to take a particle
from its rest frame to a frame
in which its energy is $E$.
As already noted above, in the case of DSR1 one finds
\begin{equation}\label{xidsr1new}
\cosh (\xi) = \frac{e^{E/E_p} - \cosh\left(m/E_p\right)}
  {\sinh\left( m/E_p \right)} ~,~~~
\sinh (\xi) = \frac{p e^{E/E_p}}
  {E_p \sinh\left(m/E_p\right)}\,\, .
\end{equation}

Relations such as these (\ref{xidsr1new}) can actually be taken as
starting point for the construction of a DSR theory based on
a nonlinear realization of the Lorentz group.
The underlying structure of a nonlinear realization of the Lorentz
group implies that relations such as (\ref{xidsr1new})
should be obtainable through a nonlinear (and, as mentioned,
physically inequivalent) redefinition of the energy-momentum variables.
In ordinary Special Relativity (governed by the linearly-realized
Lorentz group) the relations (\ref{xidsr1new}) take the form
\begin{equation}\label{xiSRnew}
\cosh (\xi) = {\epsilon \over \mu} ~,~~~
\sinh (\xi) = {\pi \over \mu}\,\, ,
\end{equation}
where I adopted the notation $\epsilon$,$\pi$ for energy-momentum
variables that transform according to ordinary Special Relativity,
and I introduced the invariant $\mu \equiv \sqrt{\epsilon^2-\pi^2}$.
The role played in DSR1 by a nonlinear realization of the
Lorentz group implies that it should be possible to introduce
throughout the theory some special combinations of the
DSR1-physical energy-momentum variables $E$,$p$
that transform instead linearly under the DSR1 boosts.
By comparison of (\ref{xidsr1new}) and ({\ref{xiSRnew}) one easily
identifies these special functions of the
DSR1-physical energy-momentum variables $E$,$p$:
\begin{equation}\label{redef}
{\epsilon(E,m;E_p) \over \mu(m)} = \frac{e^{E/E_p} - \cosh\left(m/E_p\right)}
  {\sinh\left( m/E_p \right)} ~,~~~
{\pi(p,E,m;E_p) \over \mu(m)} = \frac{p e^{E/E_p}}
  {E_p \sinh\left(m/E_p\right)}\,\, .
\end{equation}
The function $\mu(m)$ is obtained from the condition
\begin{equation}\label{limmu}
\mu(m)^2 = \lim_{p \rightarrow 0~,~~E \rightarrow m}
\left[\epsilon(E,m;E_p)^2 - \pi(p,E,m;E_p)^2 \right]
~,
\end{equation}
and then the functions $\epsilon(E,m;E_p)$
and $\pi(p,E,m;E_p)$ are fully specified by (\ref{redef}).

Similar relations can be found with analogous reasoning in all
DSR theories (with underlying nonlinear realization of Lorentz
symmetry). As mentioned, these relations of the type (\ref{redef})
can provide the starting ingredient for the construction
of a DSR theory.
Taking as starting point the functions $\epsilon(E,m;E_p)$
and $\pi(p,E,m;E_p)$ (and the function $\mu(m)$ associated
to them through (\ref{limmu}))
one immediately has an (implicit) description of the
transformation rules, and the $E(p)$ dispersion relation
automatically takes the form
\begin{equation}\label{redefdisp}
\mu(m)^2 = \epsilon(E,m;E_p)^2 - \pi(p,E,m;E_p)^2
~.
\end{equation}
The deformed boost generators can then be derived from the
structure of this dispersion relation or simply using
\begin{equation}
{\cal B}_a = i \pi_a {\partial \over \partial \epsilon}
+ i \epsilon {\partial \over \partial \pi_a}
\label{boosnormalredef}
\end{equation}
(in which $\epsilon$ and $\pi_a$ are understood as functions
of the DSR-physical energy-momentum $E$,$p_a$).

\section{First results}
\subsection{Wavelength-dependent speed of
photons and threshold conditions in collisions}
As already mentioned above, the debate on DSR theories has correctly
kept in focus the issue of the predictions concerning a possible
wavelength dependence of the speed of
photons and a possible modification of
the threshold conditions for particle production in collision
processes.
In fact, these are key aspects of relativistic kinematics,
and there are some chances of obtaining encouragement from
planned experiments.

\subsection{Transformation rules (one-particle case)}
For DSR1 and DSR2 there has been explicit detailed analysis of
the laws of transformation of energy and momentum from
one inertial observer to another.
This has also allowed to develop techniques
of analysis which are very powerful and appear to be applicable
to a relatively large class of DSR theories (although no other
example of DSR theory has been analyzed in detail).

\subsection{Maximum momentum}
As already expected upon proposing the DSR framework~\cite{dsr1dsr2},
it has been fully established
that DSR theories may provide a natural framework
in which to implement the quantum-gravity idea of a maximum
momentum and/or energy.
This result is obtained kinematically.
For example in DSR1 the laws of
transformation between inertial observers~\cite{dsr1dsr2,jurekrossano}
are such that any given momentum can be maximally boosted to the
value $E_p/c$, where $E_p$ is the observer-independent
relativistic deformation scale (naturally identified, up to
a coefficient not-too-different from 1, with the Planck scale).
In DSR2 both energy and momentum can only be maximally
boosted to maximum values set by $E_p$ and $E_p/c$ respectively.

\subsection{Laws of composition of energy-momentum}
As also already expected upon proposing the DSR framework~\cite{dsr1dsr2},
it has been fully established
that DSR theories require a nonlinear law of composition of energy
and momentum. This property can be understood
in analogy with the fate of the law of composition of velocities
in going from Galilei Relativity to Einstein's Special Relativity.
In Galilei Relativity there is no absolute velocity scale and,
as a consequence, the law of composition of velocities could not
be anything else but linear. A linear law of composition of velocities
is of course incompatible with the existence of a maximum velocity.
Special Relativity required a nonlinear law of composition of velocities.
Just like the Galileian linear law of composition of velocities
must be rejected upon introducing an absolute velocity scale,
the special-relativistic law that composes linearly the energies and
momenta of particles
must be rejected~\cite{dsr1dsr2} upon introducing an absolute
energy/momentum scale.

For DSR1 and DSR2 acceptable laws of composition of energy-momentum
(laws with the needed covariance properties under DSR transformations)
have been constructed.

\subsection{Deformed Klein-Gordon/Dirac/Maxwell equations}
The first steps have recently been taken toward the construction
of theories that are consistent with DSR kinematics.
This programme has taken has starting point the structure
of the deformation of the
Klein-Gordon, Dirac and Maxwell equations in energy-momentum space.

The Klein-Gordon equation in energy-momentum space is basically
a direct reflection of the dispersion relation, so it can be
constructed straightforwardly in DSR theories which have been
analyzed at the level of the dispersion relation, such
as DSR1 and DSR2.

In work (that progressed in parallel and interconnectedly) by
Arzano and myself~\cite{dsrdirac} and
by Ahluwalia and Kirchbach~\cite{dharamdsrdirac}
the structure of the Dirac equation for DSR1 and DSR2 in
energy-momentum space has been established.
In particular for DSR1
this deformed Dirac equation can be conveniently
written as
\begin{equation}\label{eqx}
  \left(\gamma^{\mu} {\cal D}_{\mu}(E,p,m;E_p)-I\right)
  \psi(\overrightarrow{p})=0
\end{equation}
where
\begin{equation}\label{eqy}
 {\cal D}_{0} = \frac{e^{E/E_p}-\cosh\left(m/E_p \right)}
 {\sinh\left(m/E_p \right)} ~,
\end{equation}
\begin{equation}\label{diki}
 {\cal D}_{a} = {p_a \over p} \frac{\left(2
e^{E/E_p}\left[\cosh\left(E/E_p \right)
 -\cosh\left(m/E_p \right)
\right]\right)^{\frac{1}{2}}}{\sinh\left(m/E_p \right)}
~,
\end{equation}
and the $\gamma^{\mu}$ are the familiar ``$\gamma$ matrices".

There appears to be no in-principle problem in writing
analogously a Maxwell equation, but this has not yet
been done.

\section{Open problems for DSR}

\subsection{How should one describe macroscopic bodies in DSR?}
One key issue for the DSR research programme is the one concerning the
description of macroscopic bodies. Planck-scale deformations of the
dispersion relation are clearly admissable for microscopic particles,
since we produce/observe these particles with energies that are
much smaller than $E_p$ and therefore the predicted new effects
are small enough to comply with present experimental limits.
Instead macroscopic bodies typically have energies largely in
excess of $E_p$, and the assumption of a
Planck-scale deformation of the
dispersion relation is in clear conflict with observations
as ordinary as the motion of planets in the solar system
and the game of soccer.

It has been clear from the beginning~\cite{dsr1dsr2}
that the DSR framework automatically includes structures
that may be useful in settling satisfactorily this issue,
but a full satisfactory understanding/description is still missing.

\subsection{What is the DSR observer?}
An issue which is possibly related with the above-mentioned issue
concerning macroscopic bodies is the one that concerns the
description of DSR observers.
It appears likely that the theory will require the obsever
to be a macroscopic system, to which the Planck-scale deformation
does not apply. Since we lack a description  of macroscopic bodies,
we are also still lacking a genuine understanding of DSR obsevers.

It also appears that one should think of these observers as equipped
with a large variety of probes. Since the speed of photons is
(in generic DSR theories, not in DSR2) energy dependent,
some of the perceptions of the observer should perhaps
depend on the type of probes the observer uses in a given context.
For example, in a context probed with high-energy probes
the observer might
experience a different type of time dilatation and leangth contraction.
But this has not yet been investigated.
Perhaps, rather than thinking of a single observer with different types
of probes, we should think of different types of observers, characterized
by the type of probes they use.

At the merely technical level these issues concerning the DSR observer
are relevant for establishing the relation between rapidity and
relative velocity among observers. Is this relation modified?
Or should we still adopt the special-relativistic relation
between rapidity and relative velocity?

\subsection{Which spacetimes are compatible with the DSR framework?}
All robust results obtained so far in the DSR framework concern
the energy-momentum sector.
It is unclear which type of spacetime picture is required by
the DSR framework. We should be prepared for a significant ``revolution"
in the description of spacetime. The introduction of the first
relativistic observer-independent scale, $c$, forced us to renounce
to the (until then unquestioned) concept of absolute time.
What should we give up for the second observer-independent scale?

It appears reasonable (though not the only plausible choice) to insist on
a spacetime picture which is consistent with the
relation $v=dE/dp$ between the velocity of a particle and its
energy-momentum. This relation holds in Galilei relativity, and it
turned out to survive the introduction of the first obsever-independent
scale in Special relativity, so it appears natural to assume that
it would survive also the advent of a second observer-independent scale.
This relation also means that the relativistic theory is consistent
with some kind of Hamiltonian dynamics: $dx/dt=dH/dp$.

If $v=dE/dp$ should indeed be enforced, the implications
for the spacetime picture could be profound.
In typical DSR theories, with their key nonlinearities, one observer, $O$,
could see two particles with different masses $m_A$ and $m_B$
moving at the same speed and following the same trajectory
(for $O$ particles $A$ and $B$ are ``near" at all times),
but the same two particles would have different velocities
according to a second observer $O'$, so they could be ``near" only
for a limited amount of time.
For the particles we are able to study/observe,
whose energies are much smaller than the Planck scale,
and for the type of (relatively small)
boosts we are able to investigate experimentally,
this effect can be safely neglected. But conceptually it has striking
implications. The possibility of the absolute statement ``particles $A$
and $B$ follow the same trajectory" would be removed from
our spacetime picture.

Perhaps this is taking us in the direction of considering spacetime
as an approximate concept, only valid within a certain class of
observations and with a certain level of approximation.
Perhaps such a picture could be implemented through
a description of spacetime in terms
of noncommutative geometry. Noticeably, there is at least one
noncommutative spacetime in which the DSR framework appears to
be applicable. This is the ``$\kappa$-Minkowski"
spacetime~\cite{majrue,kpoinap}
\begin{equation}
[x_j,t]= i \lambda x_j~,~~~[x_j,x_k] = 0
~,
\label{kmink}
\end{equation}
where the product of plane waves has the property
\begin{equation}
\left( e^{i p_m x_m} e^{i p_0 x_0} \right)
\left( e^{i k_m x_m} e^{i k_0 x_0} \right) =
e^{i (p_m +e^{\lambda p_0} k_m) x_m} e^{i (p_0+k_0) x_0}
\label{expprodlie}
\end{equation}
Eq.~(\ref{expprodlie})
clearly relfects a nonlinear law of composition of momenta and is
therefore consistent with the basic structure of DSR theories.

\subsection{Can DSR solve the GZK anomaly?}
As mentioned the DSR framework naturally leads to the prediction
of modified threshold conditions
for particle production in certain collision processes.
This can be perceived as an opportunity for DSR: in fact, in astrophysics
there has been much recent discussion~\cite{kifune,gactp,aus}
of anomalies possibly
related with such threshold conditions.
In particular, the analysis of a threshold condition is key for
the GZK limit on the observations of cosmic rays.
Cosmic rays travel toward us in the environment of the CMBR photons.
When the energy of the cosmic ray
exceeds the threshold value $5 {\cdot} 10^{19} eV$,
according to ordinary Special Relativity it should be possible for
the cosmic ray to loose energy through pion production off CMBR photons.
This should render observations of cosmic rays above $5 {\cdot} 10^{19} eV$,
the GZK limit, extremely unlikely.
Still, more than a dozen UHECRs have been reported by AGASA~\cite{AgaWat}
with nominal energies at or above $10^{20}$ eV.

While DSR theories do generically predict anomalous thresholds
(and therefore generically predict a different value of the GZK limit),
as mentioned in DSR1 and DSR2 the change in the value of the GZK limit
is very small, and could not be used to explain away the
paradoxical ultra-high-energy cosmic-ray observations.

Even looking beyond DSR1 and DSR2 it appears hard\footnote{Some
authors~\cite{leedsrnew} have given up so completely on the hope
of finding a DSR solution for
the cosmic-ray puzzle that they arrived at proposing that solutions
of the cosmic-ray puzzle would necessarily require {\underline{two}}
length/energy scales, rather than the single one available
in DSR theories.} to find a DSR theory
which would significantly affect the GZK limit.

The AGASA evidence of a cosmic-ray paradox must be considered as
preliminary. Forthcoming more accurate cosmic-ray observations~\cite{auger}
may well show us\footnote{It is not uncommon that preliminary data
generate interest in related theory subjects, and in some cases
the lessons learned through those theoretical studies outlast the possible
negative evolution of the experimental situation. This author is
familiar~\cite{bjpap} with the theory work that was motivated
by the so-called ``centauro events". It is now widely believed that
centauro events were a ``mirage", but in the process we did learn that
the formal structure of QCD allows the vacuum to be temporarily
misaligned (disoriented chiral condensates) and the RHIC collider
is conducting dedicated experiments. The cosmic-ray paradox might or
might not disappear when more accurate and reliable data become
available; however, in either case, the debate on the cosmic-ray
paradox has had the merit to push us to discover that certain
types of deviations from ordinary Lorentz symmetry are plausible
in quantum-gravity scenarios.}
that there is no cosmic-ray paradox after all.
Still, it is frustrating that one of the few cases in which ordinary
Special Relativity is being justifiably questioned appears to be
also a context in which the predictions of DSR theories do not
differ significantly from the ones of Special Relativity.

\subsection{What about causality?}
In a significant portion of DSR theories,
including DSR1, there appear to be also profound
consequences for causality.
These theories predict that ultra-high-energy photons
would travel faster than the low-energy photons which we ordinarily
observe/study (for which it is well established that they travel
at speed $c$). Actually, in DSR1 the limit in which a single particle
has infinite energy is also the limit in which the speed of that
particle is itself infinite. Of course, we cannot even contemplate
a particle with infinite energy (we can at best,
very optimistically of course,
contemplate the possibility to put all the energy on
the Universe in a single particle), but nevertheless it is natural
to aspect that these properties would have profound implications
for our understanding of causality.

The implications can be positive: for example, some authors have
used (see, {\it e.g.}, Ref.~\cite{cosmodsr})
some preliminary intuition about the new causality to construct
cosmological models based on DSR that would not require inflation.
These proposals provide an elegant reinterpretation
of ``varying-speed-of-light cosmology"~\cite{vsl1,vsl2,vsl3}:
rather than assuming an
explicit time dependence of the speed of photons (which would inevitably
lead to a breakup of Lorentz symmetry), one works within
a DSR framework {\it a la} DSR1, in which the speed of photons increases
with energy, and then observes that in the Early Universe particles
typically had very high energies and (within DSR1) could put in causal
connection regions of the cosmo which instead would be causally disconnected
according to ordinary Special Relativity.

But a full understanding of DSR causality is clearly a key objective
of this research programme. We should perhaps classify inertial observers
in classes depending on the probes they have available.
Would the observers which have ultra-high-energy photons at their disposal
be able to introduce a concept of time that is (to good approximation)
absolute?
The dilatation of the muon lifetime due
to its velocity would depend on which
probes we use to establish this lifetime?

\subsection{What about noninertial observers?}
Of course, since the DSR research programme is considering a modification
of Special Relativity, a natural next step to consider is the one
in which an analogous modification is implemented at the level
of General Relativity.
Nothing that can claim any robusteness has been obtained so far
on this interesting point.
I believe~\cite{dsrpolon}
that a key issue for such studies comes from the observation
that we might be required to attribute to the Planck scale a double role:
a role in the gravitational coupling (because of the relation between $G$
and $E_p$) and a role in the structure of spacetime (energy-momentum space).
If this intuition turns out to be correct we might have to face
significant challenges at the conceptual level. It is always very
significant when two operatively well-defined concepts turn out
to be identified (see, {\it e.g.}, the Equivalence Principle for
inertial and gravitational mass).

\section{Some ideas for the open problems}
As I was discussing the ``open problems" for the DSR research programme
I already mentioned a few ideas that may prove useful, as in the case
of inertial observers classified on the basis of the typical energies
of the probes available to them.
For some other ideas I thought it might be appropriate to have
this dedicated section.

\subsection{Macroscopic bodies severely affected by the nonlinearity
of DSR theories}
The fact that ordinary Special Relativity has an absolute velocity scale
forced to compose velocities nonlinearly. We compose velocities only
in one context: when we compare the velocity that a particle has for
a given observer $O$ with the velocity that the same particle has
for another observer $O'$ (itself moving at some velocity with
respect to $O$).
In DSR theories there is also a large-energy/small-length scale
and this of course imposes that energy-momentum be composed nonlinearly.
We clearly need to compose energy-momentum if we want to impose
some sort of conservation of ``total"~\cite{dsr1dsr2,dsrpolon}
energy-momentum in collision processes.
There is also another context in which we, in a certain sense,
compose energy-momentum: a macroscopic body is ``made of" a large
number of microscopic particles, and its own energy-momentum is
obtained ``composing" the energy-momentum of the microscopic particles
it is made of and taking into account binding energies.

When we compose energy-momentum in the context of conservation laws
for particle-production processes we are genuinely dealing
with pure kinematics. But when we describe a macroscopic body in terms
of the microscopic particles it is made of it is necessary to
consider also dynamics. Perhaps this description of macroscopic
bodies (for which we are presently not ready) will provide the
answers we are seeking. One can even speculate that through the mechanism
that allows a macroscopic body to be formed out of many microscopic
particles the DSR effects might be screened, just like
in quantum mechanics the quantum properties of
microscopic particles are screened in cases in which these microscopic
particles are ``put together" into a macroscopic body.
The analogy with quantum mechanics finds also some intuitive
(though not yet robust)
support in the fact that the Planck scale does involve Plack's
constant $\hbar$.

\subsection{Macroscopic bodies made of particles with minimum
wavelength}
Perhaps instead the clarification of the ``macroscopic-body problem"
will come from other mechanisms. Perhaps DSR should be used to describe
a minimum wavelength rather than a maximum momentum.
This would at least provide an acceptable description of beams
of particles. A laser
beam of photons has much larger momentum that the momentum
of its composing photons, but both the beam and each of the
composing particles have the same wavelength.
A deformed wavelength/frequency dispersion relation is acceptable
for both the beam and the single photon, while a deformed energy/momentum
dispersion relation cannot be applicable to the macroscopic beam.

A beam of particles is much simpler than a macroscopic body (in
which microscopic particles are bound), but
also for a macroscopic body
a deformed wavelength/frequency relation
coudl differ from a deformed energy/momentum relation
in a significant way.

\subsection{Macroscopic bodies a la DSR3}
Perhaps the solution of the ``macroscopic-body problem"
is even simpler. In DSR theories of type DSR3 one can easily
obtain that the effect is automatically confined to particles
with small mass.
In particular, a DSR deformation of Special Relativity based on
a relation of the type (\ref{dispgood}) could be applied
directly to macroscopic bodies, since it leads to
vanishingly small effects in the large-mass limit.
Moreover, in spite of the $E^2/(m E_p)$ dependence,
the relation (\ref{dispgood}) is structured in such a way that
it also leads to
the absence of any deformation for massless particles.
The DSR effects suggested by the DSR3-type
relation (\ref{dispgood}) are significant only for particles with
small mass $m$ (much smaller than $E_p/c^2$)
and only when these particles have energies in
the neighborhood of $\sqrt{m E_p}$.

\subsection{GZK anomaly a la DSR3}
The DSR3-type theories which I proposed to consider in Ref.~\cite{dsrgzk}
might also provide a straightforward solution for the cosmic-ray paradox.
As shown in Ref.~\cite{dsrgzk}, contrary to DSR1 and DSR2,
these DSR3-type theories can lead to a significant shift of the GZK
limit.

\subsection{GZK anomaly as a reflection of the properties of
composite particles}
The DSR framework naturally leads~\cite{dsr1dsr2,dsrpolon}
to the concept of a fundamental building block for particles.
In fact, DSR usually requires different kinematical properties
for composite particles (up to macroscopic bodies)
and fundamental particles.
For the first time in the history of physics there might be a meaningful
(rather than phylosophical) issue concerning the identification of
fundamental particles. Which particles are fundamental?

The GZK limit for cosmic-ray observations crucially depends~\cite{kifune,gactp}
on the analysis of photopion production $p + \gamma \rightarrow p + \pi$
(incoming proton and photon, outgoing proton and pion).
Should we apply the charateristic dispersion relation
of a given DSR to pions and protons?
Or should we describe pions and protons as some sort of DSR
bound state of quarks (which, as a DSR multiparticle body,
might obey a different dispersion relation)?
Should we even think of quarks and photons as fundamental?
An ordinary photon is huge with respect to the Planck length!

The DSR analysis of the GZK limit might therefore be highly nontrivial.
It might be~\cite{dsrnature} connected with the problematic
description of bound systems and macroscopic bodies.

Incidentally, let me observe that
this perspective on the cosmic-ray paradox
is also connected with the questions about a DSR spacetime,
mentioned in Subsection~6.3. If we allow different particles
(protons, pions, photons...) to obey different kinematic laws,
then the scenario in which spacetime is a derived approximate
concept appears to become inevitable. One could abstract a spacetime
in contexts (the only contexts to which we presently have access)
in which all particles have energy/momentum much smaller than
the Planck scale, but in contexts in which particles with energy/momentum
comparable to the Planck scale are available (like the early Universe)
it would not even be possible to introduce a good approximate concept
of spacetime.

\section{Comparison with previous literature}
The DSR literature has grown rapidly in this first two years
of existence. Many robust results have been obtained.
Naturally,
some incorrect descriptions have also surfaced in the DSR literature.
In particular, there are very different views on the connection
between DSR theories and other ideas for departures from
Special Relativity which have also been considered in the literature.

An analysis of research developments over the last few years
shows that the true root of the DSR proposal put forward
in Refs.~\cite{dsr1dsr2} is in the previous studies~\cite{grbgac,gampul}
which were
based on the idea of spacetime-foam-induced deformed dispersion relations.
Those spacetime-foam scenarios require that Lorentz symmetry be broken,
that there be a foamy spacetime background and an associated
preferred class of inertial observers.
At some point this author came to ask the question: if a quantum-gravity
scenario is shown to lead to a Planck-scale-deformed dispersion
relation, does this automatically imply that Lorentz symmetry is
broken and there is a preferred class of inertial observers?
The answer turned out to be negative.
It turned out to be possible to describe kinematics in terms
of a new class of relativistic theories, the DSR theories,
in which a deformed dispersion relation is implemented by deforming
(rather than breaking) Lorentz symmetry and without any preferred
class of inertial observers.

Some of the ingredients of a DSR theory are also present in
other new-physics scenarios, most notably the intuition
that the Planck scale might be one or another kind of
fundamental scale. But in DSR the Planck scale is introduced
as a specific type of fundamental scale, and the DSR
framework prescribes certain connections between the Planck
scale and other physical entities. These characteristic features
of DSR were summarized in Section~3:

\noindent
{\bf (i)} DSR poses a precise condition on the
second observer independent scale. It should be a fundamental
scale in the sense of $c$, and {\underline{not}} in the
sense we presently attribute to $\hbar$.

\noindent
{\bf (ii)} Since a key condition is that the second observer-indepedent
scale be treatable in complete analogy with what we presently do
with $c$, the main focus of DSR research, the first structure that
must be explored in a DSR proposal, is the description of
transformation laws between inertial observers.
And since the ultimate goal is the description of
actual observations/data, one cannot be satisfied with infinitesimal
transformations between observers. Boosts of arbitrary magnitude
must be considered.

\noindent
{\bf (iii)} The second observer-independent scale should become
significant in the ultra-high-energy/ultra-small-wavelength regime.

\subsection{Works on minimum wavelength, maximum acceleration,
and similar concepts}
The idea that besides the maximum velicity $c$ and minimum (non-zero)
angular momentum $\hbar$ there might be other similar constraints,
like a minimum wavelength and a maximum acceleration,
has been explored through many views/scenarios in the literature.
Although the DSR framework does not necessarily lead to this type
of constraint (the second {\underline{relativistic}} observer-independent
scale can be introduced in other ways), much of the work on DSR
has been focusing on the possibility of
a maximum-momentum/minimum-wavelength (and maximum energy, in some cases),
and in this respect it is naturally connected with this previous literature.
The connection however stops at the level of this type of constraint.
The DSR requirements that the second observer-independent scale
be {\underline{relativistic}} ({\it i.e.} similar to $c$,
not to $\hbar$, see Subsection~3.1),
and that this should be reflected in deformed transformation rules
between inertial observers, and that the deformation should affect
primarily the high-energy regime, were not put in focus
before work in the DSR framework.

In Refs.~\cite{kempmang,dadebro}
deformations of the Heisenberg algebra (phase-space deformations)
were explored as a way to obtain the existence of a minimum wavelength.
Basically, one would find a minimum wavelength because of
a deformed relation between momentum and wavelength.
Momentum would still be allowed to go up to infinity,
but the relation between momentum and wavelength would be such
that infinite momentum would still correspond to a finite minimum
value of wavelength.
Since in this scenario the minimum wavelength is not the result
of deformed transformation laws between inertial
observers\footnote{The laws of transformation of energy-momentum
(and spacetime coordinates)
were not even investigated in Refs.~\cite{kempmang,dadebro}.},
there is at present no connection with DSR research,
although it might be intersting to seek a scheme in which
this mechanism for minimum wavelength is implemented {\it a la}
DSR.
Similar remarks apply to other ``minimum wavelength", ``minimum length",
``minimum length uncertainty" proposals.

Refs.~\cite{maxA1,maxA2} elaborated the idea of
a maximum proper acceleration (from which a
minimum length uncertainty eventually emerged),
by considering a
phase-space approach to quantum geometry.
A deformation of the rules of transformation between
inertial observers
was not even considered, and, as a consequence, it is difficult
to make any assumption about the nature of the maximum-acceleration
constraint (should it be {\it a la} $c$? or {\it a la} $\hbar$?).
It would be interesting to attempt to implement {\it a la} DSR
this idea of a maximum acceleration.

\subsection{Works on Scale Relativity}
Scale Relativity is a proposal of revision of the principles of
relativity that
originated in works by Nottale~\cite{nottale}
and was further developed through work also
by other authors
(see, {\it e.g.}, Refs.~\cite{pronott1,pronott2}).
It is based on the idea that the relativity principles should
involve, in addition to the usual transformation between
inertial observers (primarily characterized by the relative
orientation of the axes of the observers and the relative
velocity of the observers),
also a certain type of scale transformations.
This is clearly a ``beyond special relativity" proposal
that is complementary to the DSR proposal.
While DSR postulates that the ordinary transformations
between observers should be deformed in order to accomodate
a second relativistic observer-independent scale,
the Scale-Relativity research programme does not necessarily
question the form of the ordinary rotations+boosts transformations,
but rather postulates that in addition to these transformations
one should consider an additional class of (scale) transformations.

While as originally formulated (and presently studied) the DSR proposal
and the Scale-Relativity proposal are complementary and
alternative, it would be interesting to explore the possibility of
merging these two proposals (if not in a rigorous sense, at least
in the sense of combining some of the ingredients of both).
As mentioned in discussing here some of the key open issues
for DSR research, the deformed energy-momentum transformations laws
of typical DSR schemes could invite us to classify observers
according to the energy of the probes they are using,
and this might force us to consider a corresponding class of
scale transformations.

Similarly, it is not unconcievable that the addition of scale
transformations postulated by the Scale-Relativity programme
might lead to some consistency requirement concerning the
action of rotations and boosts in the energy-momentum sector.
The recent interest in the DSR framework should motivate a
reanalysis of the Scale-Relativity framework in which the
focus be placed on the structure of finite rotation and boost
transformations, and such that the possible emergence of
an observer-independent scale in the DSR sense could be uncovered.
This could lead to a Scale-Relativity extension of the DSR
framework, rather than of the ordinary Special-Relativity framework.

\subsection{Works on Fock spacetime}
It has been noticed (already in the paper~\cite{leedsr} that proposed DSR2)
that the DSR2 energy-momentum transformation rules are formally
related to some spacetime transformation rules which were written
down long ago by Fock~\cite{fock}.
It should be stressed that Fock was not guided by plans to revise
the Special-Relativity postulates (and of course he had no intention
of introducing a second observer-independent scale).
Fock was rather interested~\cite{fock} in establishing the precise
relation between the conceptual structure of Special Relativity
and the nature of the Special-Relativity transformation laws.
The type of question Fock was intending to investigate is:
if I remove one or another conceptual element of Special Relativity
which corresponding generalization of Lorentz spacetime transformations
becomes allowed?
In one case (removing a corresponding conceptual ingredient of
Special Relativity) Fock found a one-parameter family of
spacetime transformation laws which turns out to coincide formally
with the DSR2 energy-momentum transformation laws.

Fock's motivation was not the one of introducing a second
observer-independent scale, and the formulas he eventually stumbled upon
require a {\underline{large distance}} scale. The corresponding
physical effects would be most significant at low energies,
contrary to one of the primary conceptual ingredients
that characterize the DSR framework (and in clear conflict with
everyday observations).
The Fock result (which was not even the proposal of a physical
theory, it was just an observation on the logical structure
of ordinary Special Relativity) is physically completely
different from the DSR2 proposal in spite of the amusing
similarities of the mathematical formulas.
Attributing to Fock the DSR2 proposal of Ref.~\cite{leedsr}
would be a bit like attributing to Bardeen-Cooper-Schrieffer
(and their study of superconductivity)
the proposal of the Glashow-Weiberg-Salam Standard Model
of particle physics.

\subsection{Works on $\kappa$-Poincar\'{e} Hopf algebras}
As mentioned, the research programme which has closer connection
to the DSR proposal is the one in which quantum gravity
was advocated in motivating
the emergence of deformed dispersion relations.
Before the DSR proposal, authors would automatically assume that
these deformed dispersion relations should
require that Lorentz symmetry be broken,
with the emergence of an associated
preferred class of inertial observers.
DSR was proposed primarily with the intent/objective of
observing that this might not be true in general:
a Planck-scale deformed dispersion relation can be adopted
without the support of a preferred class of inertial observers,
at the ``cost" of a corresponding deformation of
the laws of transformation between inertial observers.

Deformed dispersion relations had also been encountered in
investigations of
the $\kappa$-Poincar\'{e} Hopf algebras~\cite{kpoinfirst,rueggnew,kpoinap}.
The mathematics research line of ``quantum groups" had encountered
serious resistence in obtaining a quantum deformation of
the Poincar\'{e} algebra. The $\kappa$-Poincar\'{e} research line emerged
out of the idea of circumventing these difficulties
by first achieving a quantum deformation of the symmetry algebra
of deSitter space, and then taking a cleaver limiting procedure,
which would lead to candidate ``quantum" versions of the Poincar\'{e}
algebra.
Many such $\kappa$-Poincar\'{e} Hopf algebras
have been considered~\cite{kpoinap}.
They were not born out of the objective of introducing
a second relativistic observer-independent scale, and actually
the question about observer-independent scales could not even
be properly formulated in the $\kappa$-Poincar\'{e} framework,
since results concerning the exponentiation of
Lorentz-like $\kappa$-Poincar\'{e}
algebra elements had been found~\cite{rueggnew} to
have puzzling properties (these exponentiations of algebra elements
could not be seen as elements of a group, but only as elements
of a ``quasigroup" in the sense of Batalin~\cite{batalin}).
In a sense, from the technical side,
the opportunity for the DSR proposal came out of the
realization~\cite{dsr1dsr2} that these results about problems
with the exponentiation of the Lorentz-like
elements $\kappa$-Poincar\'{e} Hopf algebras were not present in
all $\kappa$-Poincar\'{e} Hopf algebras. The results of Ref.~\cite{rueggnew}
had led to the expectation that one would encounter these problems
in all $\kappa$-Poincar\'{e} Hopf algebras, but actually the rotation/boost
transformations of the DSR1 proposal can be cast in the framework
of a particular $\kappa$-Poincar\'{e} Hopf algebra
(the one called ``bicrossproduct basis" in the mathematics literature),
and in that particular $\kappa$-Poincar\'{e} Hopf algebra
one encounters no problems concerning the exponentiation of
algebra elements, thereby providing a counter-example
for the scenario which had emerged in Ref.~\cite{rueggnew}.

The fact that the concept of relativistic transformations between
inertial observers was very far from the objectives of
work on $\kappa$-Poincar\'{e} Hopf algebras is also reflected in the
structure of the laws of composition of energy-momentum
which had been adopted in the relevant literature (see, {\it e.g.},
Ref.~\cite{dsrlukie} and references therein).
According to these laws a particle-producing
collision process $a+b \rightarrow c+d$
would lead to a ``energy-momentum-conservation" condition of
the type $(p_a \dot{+} p_b)^\mu = (p_c \dot{+} p_d)^\mu$,
where $\dot{+}$ is a deformed and nonsymmetric (nonabelian sum)
law of composition.
While the fact that the composition law does not apply symmetrically
to the intervening particles had led to some concern, there was no
concern (it was not even noticed) about the fact that
such laws are not truly covariant under action through
the $\kappa$-Poincar\'{e} algebra elements.
Acting with $\kappa$-Poincar\'{e} algebra elements on
$(p_a \dot{+} p_b)^\mu$ on does obtain the transformed
$(p_a' \dot{+} p_b')^\mu$,
but unfortunately it is easy to verify~\cite{areaNEWpap}
that the condition $(p_a \dot{+} p_b)^\mu = (p_c \dot{+} p_d)^\mu$
is incompatible with the
condition $(p_a' \dot{+} p_b')^\mu = (p_c' \dot{+} p_d')^\mu$.
In the new perspective of the DSR proposal one would describe
these $\kappa$-Poincar\'{e} laws as inconsistent with
the Relativity Principle.
While in certain DSR schems, like DSR1, the one-particle-sector
transformation rules can be formulated in terms of
({\underline{exponentiated}}) $\kappa$-Poincar\'{e} boost generators,
this problem about composition of momentum actually poses a
serious obstacle for the interpretation {\it a posteriori}
(in light of the DSR proposal) of at least
some $\kappa$-Poincar\'{e} algebras as mathematical basis for
a DSR physical theory.

Setting aside these problems in the multiparticle sector,
it appears proper to perceive the relation between
certain DSR theories
and work on corresponding $\kappa$-Poincar\'{e} Hopf algebras
in the same way in which we perceive the relation
between Einstein's physical theory of Special Relativity
and preceding works by FitzGerald, Lorentz and Poincar\'{e}.
Some $\kappa$-Poincar\'{e} Hopf algebras provide an important
mathematical background for certain corresponding
DSR physical theories.
In the DSR theories in which this correspondence is present
it may be appropriate to call the transformation
laws ``$\kappa$-Lorentz transfomartion laws", just like we
call Lorentz transformations the transformation laws of
Einstein's Special Relativity.

\section{Outlook}
Doubly-Special Relativity is maturing quickly, as a result
of the interest it is attracting from various research groups,
each bringing its relevant expertise to the programme.
Some key results have already been obtained. I have emphasized here
the ones that I presently perceive as most significant.
Several crucial open issues remain to be tackled.
I have attempted to list the most important of these open issues,
and I also ventured suggesting possible ``lines of attack".

DSR research also clearly must find some additional
consistency requirements. For example, it appears
plausible that not all nonlinear
realizations of the Lorentz group give rise to a physically
acceptable DSR theory. But we still lack a suitable consistency
requirement that would allow us to identify the cases
that must be disregarded.
Perhaps some condition of consistency with the emergence
of an ``acceptable spacetime picture" (whatever that means)
should be implemented.

Of course, the most important results that this research line is
awaiting are experimental results. We can be moderately optimistic
that with new observatories, such as GLAST
and the Pierre Auger Observatory, we might have a key experimental
hint within a few years. Without this type of experimental guidance
work in DSR will remain in the awkward limbo in which Einstein
would have been in trying to introduce an observer-independent velocity
scale without having available the indication that this observer-independent
velocity should be the one of the Maxwell equations and of
the Michelson-Morley experiments.
We must therefore think hard of other experimental contexts which
may be of help for the development (or falsification) of DSR.
New opportunities might materialize even on a relatively
short time scale. In these closing remarks I want to contemplate
an example of such an opportunity which might materialize
in the not-so-distant future.
I expect that the type of time-of-arrival-difference
studies that GLAST will be conducting~\cite{glast}
should somehow be doable even with particles of much higher energies.
For example, it is sometimes argued that the mechanism
that produces ultra-high-energy
cosmic rays is the same mechanism that gives rise to gamma-ray bursts.
If gamma-ray bursters also emit (roughly simultaneously with the
gamma-ray burst) particles of extremely high energies,
we could exploit the fact that,
over a time of travel of $10^{17}s$,
even with a dispersion relation that is deformed only
at the $E_p^{-2}$ level
(quadratic Planck-scale suppression), a particle with $E\sim 10^{19} eV$
can acquire a time-arrival difference with respect to particles
of $E\sim 10^6 eV$ (particles which arrive with the main
burst) which is of the order of $0.1 s$, and possibly observable.


\nonumsection{References}

\end{document}